\documentclass[11pt]{article}
\usepackage[margin=1in]{geometry}
\usepackage{graphicx}
\usepackage{booktabs}
\usepackage{amsmath}
\usepackage{amssymb}
\usepackage[hidelinks]{hyperref}
\usepackage{xcolor}

\title{Measuring Alignment With Reader Highlights\\
Net of Position and Length}
\author{
  Kazuki Nakayashiki \quad Keisuke Watanabe \\[3pt]
  Glasp Inc. \\
  \texttt{kazuki@glasp.co} \quad \texttt{kei@glasp.co}
}
\date{}

\begin{document}
\maketitle

\begin{abstract}
Context compression discards most of a document before a language model reads it, and is normally
evaluated by downstream task accuracy --- which makes another model the judge of what mattered.
Naturalistic social highlighting offers a non-circular reference: many people independently
marking passages on the same page. But the obvious metric, the fraction of crowd-marked sentences a
compressor keeps, is confounded twice: crowd marks are front-loaded and crowd-marked sentences are
longer, so any method favouring early or long sentences scores well regardless of readers. We
remove both by matching each marked sentence against unmarked sentences of the same document at
equal relative depth and equal within-document length rank, and we calibrate every estimator on
synthetic nulls built from position and length alone --- a step that matters, since depth-only
stratification returns a false positive on 20--36\% of nulls containing no effect. On 120 web
documents ($\geq$12 independent readers each), a language-model importance ranking keeps 38.4\% of
crowd-marked sentences against 19.9\% of their matched neighbours: an enrichment of $+0.196$
[$+0.148$, $+0.239$], at $p = 0.0005$ under an exact randomization test that assumes nothing about
clustering, and replicated cross-vendor. Naive truncation, whose keep rule \emph{is} position,
correctly falls to $+0.003$. To give the number a scale: scored identically, on the same budget,
against a crowd label recomputed to exclude them, \textbf{a single human reader reaches $+0.182$}
--- indistinguishable from GPT-5.4 ($+0.002$ [$-0.081$, $+0.088$]) and below Claude Opus 5. Classical methods are not null --- Luhn's 1958 heuristic reaches $+0.088$ --- so reader selection
is partly recoverable by counting words; conditioning additionally on lexical centrality removes
only $0.010$, so the agreement is not centrality. We also report that a claim in our own
prior work does not reproduce on this corpus.
\end{abstract}

\section{Introduction}

Long-context pipelines routinely discard 80\% of a document before the model sees it. Whether
a compressor keeps ``what matters'' is normally assessed by downstream task accuracy, which is
circular in a specific way: the judge of what mattered is another model.

Social highlighting offers a non-circular reference. When many people independently read the
same public web page and mark passages for their own later use, the marks are an unprompted,
unpaid record of what readers chose to keep. The natural metric follows:
\[
F(c, r) = \frac{|\,\text{crowd-marked sentences} \cap \text{kept by } c \text{ at ratio } r\,|}
               {|\,\text{crowd-marked sentences}\,|}.
\]

We are careful about what this measures. We do \emph{not} claim highlights mark what is
objectively important; the educational-psychology literature contradicts that, and our argument
never needs it. We claim only that highlights record what readers chose to keep, which is true
by construction of the measurement.

$F$ taken naively is uninformative, because it is confounded twice. Crowd marks are
front-loaded and crowd-marked sentences are longer, so truncation and a length-ranking selector
both score well on $F$ while carrying no information about reader choice: conditioned on depth,
truncation's apparent preference is $-0.001$, exactly as it must be, since its keep rule
\emph{is} depth.

The question worth asking is what survives removal of both confounds. The surviving signal is partly recoverable by
cheap lexical methods --- Luhn's 1958 heuristic reaches $+0.088$ --- and roughly twice as
recoverable by a language-model importance ranking, which replicates across two vendors and retains sign and significance under every
perturbation we applied.

\paragraph{Contributions.}
(1) \textbf{A measured false-positive rate for the confound
control itself.} Matching human-marked salience on position and length is standard practice ---
Danescu-Niculescu-Mizil et al.~\cite{danescu2012} did it fourteen years ago and Tan et
al.~\cite{tan2018} match highlights against non-highlighted sentences within speaker --- and we
claim no novelty for it. What no paper in this area reports is the error rate of its own control.
Measured on synthetic nulls, depth-only stratification --- the control one reaches for first ---
returns a false positive on 20--36\% of nulls containing no effect, against a nominal 2.5\%, and
the calibration is itself sensitive to how strong the synthetic confound is made. (2) The main result: a language model keeps 38.4\% of
crowd-marked sentences against 19.9\% of their matched neighbours, replicated cross-vendor and
significant under an assumption-free randomization test. (3) A post-matching balance table that
bounds what any single surface feature could contribute, rather than asserting that none does.
(4) Evidence that prompt framing changes the result by a factor of two, in a non-obvious
direction. (5) Classical extractive methods are not null here: a 1958
heuristic recovers roughly half the effect, so the language model doubles rather than
categorically exceeds cheap lexical selection. (6) A de-identified per-sentence reproduction
artifact, and the main table, calibration, inference, baseline and centrality-control numbers as
runnable code. Claims that adversarial review overturned during this work are listed in
Appendix~A rather than folded into the contributions.

\section{Related work}

\paragraph{Position and length bias.} That extractive summarizers exploit lead position is
well established. Kedzie et al.~\cite{kedzie2018} report that ``sentence position bias
dominates the learning signal'' and ask whether models ``identify important content or just
find the start of the document''; Zhong et al.~\cite{zhong2019} reach 40.08 ROUGE-1 from
positional embeddings alone against a 41.31 full model. Grenander et al.~\cite{grenander2019}
and Jung et al.~\cite{jung2019} establish domain dependence. Length is equally old: Sun et
al.~\cite{sun2019} show ROUGE F1 confounds content selection with output length, and Kupiec et
al.~\cite{kupiec1995} used position and sentence length \emph{as the baseline} in 1995, finding
that content features slightly hurt. Matching on these two for human-marked salience is
likewise established: Danescu-Niculescu-Mizil et al.~\cite{danescu2012} matched on position and
length to license a phrasing claim, and Tan et al.~\cite{tan2018} match highlights against
non-highlighted sentences of similar length within speaker. Our contribution is not the estimator
but its calibration --- we measure what that control's error rate actually is.

\paragraph{Human reading attention.} Hollenstein and Beinborn~\cite{hollenstein2021} report
that ``both token length and frequency are strongly correlated with relative importance'';
Morger et al.~\cite{morger2022} name controlling for them as future work.
Bestgen~\cite{bestgen2021} is the sharpest warning: length and position alone reach MAE 4.219
in gaze prediction against a winning full-feature system at 3.813. Surface features dominate
human reading attention, which is precisely why an alignment result must be reported net of
them.

\paragraph{Rationale plausibility.} The closest analogue is agreement between model-extracted
and human rationales. ERASER~\cite{deyoung2020} supplies random baselines for faithfulness but
not plausibility; Atanasova et al.~\cite{atanasova2020} include a random baseline for human
agreement. Neither examines position or length. Jacovi and Goldberg~\cite{jacovi2020} give the
motivation: a plausible but unfaithful interpretation is the worst case.

\paragraph{Compression evaluation.} LLMLingua~\cite{jiang2023}, LLMLingua-2~\cite{pan2024} and
Selective Context~\cite{li2023} evaluate on downstream proxies. \L{}ajewska et
al.~\cite{lajewska2025} evaluate information preservation without human annotators.
ECoRAG~\cite{ecorag2025} does align compression against human-annotated evidentiality, but
those are question-conditioned crowdworker evidence labels rather than unprompted reading
behaviour. HighRES~\cite{hardy2019} uses elicited highlights as summarization ground truth,
under an explicit instruction and a word budget --- conditions naturalistic marks do not meet.

\paragraph{The live state of the art.} Closest to this work in question and furthest in substrate,
``What Matters to an LLM?''~\cite{whatmatters2026} extends Trienes et al.\ and finds that language
models converge on importance patterns sharply different from pre-LLM baselines --- using no human
highlight data. Our Luhn result is a partial complement: a pre-LLM word-frequency heuristic
recovers half of what a model recovers against \emph{reader} behaviour, so the divergence they
report at the level of model-versus-baseline agreement does not carry over intact to agreement
with readers.

\paragraph{A result we appear to contradict.} Trienes et al.~\cite{trienes2025} report that LLM
salience ``only weakly correlates with human perceptions of information salience'' across 13
models. We report a substantial correlation, net of two confounds they do not control. Both
cannot be casually true, and we do not think the difference is that one side is simply wrong
about the models. Three differences are candidates and we can rule on only one. Their human
signal is 18 recruited annotators rating Questions-Under-Discussion on a Likert scale; ours is
unprompted marking by ordinary readers on pages they chose to read --- a different construct,
and plausibly an easier one for a model to hit. Their systems are summarizers; ours is an
explicit ranking prompt, and \S\ref{sec:prompt} shows prompt framing alone moves our own numbers
by a factor of two. And their comparison is not net of position or length, which by
Table~\ref{tab:calib} can move an uncontrolled estimate substantially. We flag this as
unresolved rather than claim it favours us.

\section{Data}

Public web documents from a social highlighting platform, each marked by at least twelve
independent readers. Sentences are labelled crowd-marked if they fall in the top 15\% by mark
count with at least two marks. The evaluation set is \textbf{120 documents across 78 domains}.

\paragraph{What the corpus is not.} A convenience sample of one platform's users; no claim of
representativeness. The evaluation set is entirely Latin-script (mean Latin-letter fraction
0.9999), but Latin script does not imply English: 119 documents are English and one is
Portuguese-language legislation, which we leave in and flag rather than quietly drop. Documents
survive a Readability extraction, skewing toward static HTML. Sentences outside 30--400
characters are dropped and documents are capped at 300 sentences, so compressors operate on a
length-censored subsequence. Marks are anchored by substring containment, so a reader
highlighting a paragraph marks every sentence in it, and marked sentences are consequently
strongly contiguous: 54.9\% have a marked immediate neighbour against 25.3\% under size-matched
random placement, giving 1,378 runs over 2,101 marked sentences. The independent unit is
therefore closer to a run of marks than to a sentence,
and within-document precision is correspondingly overstated. Contiguity also depletes the
nearest-in-depth comparator pool, which is part of why matching discards as much as it does.
Vendor-owned domains are excluded by rule.

\paragraph{A correction to our own prior work.} Earlier work of ours~\cite{ownprior} characterised this
population using an upper bound on reader count, a per-domain cap and a sampler biased toward
thinly-read documents, and described the usable universe as physically limited to roughly 130
documents. That was an artifact of those gates: removing the ceiling yields 966 candidates and
the 120/78 set used here.

\section{Method}

\subsection{Estimand}

For compressor $c$, let $S_d$ be the crowd-marked sentences of document $d$ and $M(i)$ the
unmarked sentences of $d$ within $0.05$ in relative depth and $0.05$ in within-document length
rank. Then
\[
E_d = \frac{1}{|S_d|}\sum_{i \in S_d}\Big(\mathbb{1}[i \in K_c] - \frac{1}{|M(i)|}\sum_{j \in M(i)} \mathbb{1}[j \in K_c]\Big),
\]
aggregated across documents with a bootstrap clustered by domain.

\paragraph{What matching costs.} Matching is expensive in data: \textbf{1,304 of 2,101 marked
sentences are matched and 797 (37.9\%) are discarded} for having no admissible partner,
\textbf{15 of 120 documents contribute no pairs at all}, and comparator pools are small --- the
median is 2. Discards
are systematic, though not in the direction that would flatter us: discarded marks are
\emph{not} more front-loaded than retained ones; they come from smaller documents and are
length-extreme. Tightening the tolerance to 0.03 raises the discard rate sharply, which is why
we report the whole tolerance grid rather than a preferred setting. Because pools are small and
marks are contiguous, we treat the within-document precision as optimistic and rely on the
domain-clustered interval.

\subsection{Calibration}

An estimator that claims to remove a confound must be shown to remove it. We generate nulls
over the real corpus geometry in which the keep set is produced from position and length only,
with marking playing no role, so the true conditional enrichment is exactly zero
(Table~\ref{tab:calib}, 200 replications).

\begin{table}[t]
\centering
\begin{tabular}{lrrrr}
\toprule
& \multicolumn{4}{c}{false-positive rate under a null with no human effect} \\
\cmidrule(lr){2-5}
estimator & strong confound & centrality-like & LLM-like & no confound \\
\midrule
depth deciles                          & 36.5\% & 28.0\% & 20.5\% & 5.0\% \\
length terciles                        & 100\%  & 4.5\%  & 16.0\% & 4.0\% \\
depth quintile $\times$ length tercile & 13.0\% & 5.5\%  & 4.5\%  & 6.0\% \\
\textbf{matched (primary)}             & \textbf{4.5\%} & \textbf{5.5\%} & \textbf{7.0\%} & \textbf{5.5\%} \\
\bottomrule
\end{tabular}
\caption{Nominal rate is 2.5\%. Keep sets are generated from position and length only, over the
real corpus geometry, so the true conditional enrichment is exactly zero. Columns vary the
strength of the synthetic confound: ``strong'' is a hand-set front bias; the middle two are
tuned so the synthetic keep rule reproduces a real arm's measured dependence on depth and
length; the last has no confound at all. \textbf{The calibration is itself sensitive to this
choice}, so all four are shown. Two readings matter. Depth-only stratification degrades as the
confound it fails to remove gets stronger --- 5.0\% with no confound, 20.5--36.5\% with one ---
which is exactly the signature of a control that does not control. And the 100\% for length-only
terciles is an artifact of an unrealistically strong null, falling to 4.0--16\% otherwise; we
withdraw the wider range we first reported. Matching is 4.5--7.0\% throughout. All estimators
detect planted effects of $+0.25$ and $+0.50$ at 100\%, so this is specificity, not power.
Nulls reproduce with \texttt{robustness.py}; the planted-effect check is
\texttt{control-calibration.py}, run under the strong-confound geometry only.}
\label{tab:calib}
\end{table}

\label{sec:calib}
\paragraph{On the decision threshold.} We fixed a threshold of $+0.054$ --- the null's 97.5th
percentile --- in version control before running the estimator on the final keep sets. We
initially described this as a confound floor. \textbf{That description was wrong}: a null with
\emph{zero} position and length dependence yields $+0.060$, so the quantity is approximately
two standard errors of the estimator at $n{=}120$, not a bias correction. We report it as a
magnitude reference and rely on the intervals.

\section{Results}

\subsection{Main result}

\begin{table}[t]
\centering
\begin{tabular}{lrrr}
\toprule
arm & raw & matched (primary) & clean control \\
\midrule
\texttt{random} (calibration)   & $-0.005$ & $+0.009$ \ [$-0.045$, $+0.060$] & $-0.013$ \\
\texttt{lead} (calibration)     & $+0.097$ & $+0.003$ \ [$-0.012$, $+0.018$] & $+0.007$ \\
\texttt{length}                 & $+0.064$ & $+0.015$ \ [$-0.005$, $+0.043$] & $+0.003$ \\
\texttt{centrality}             & $+0.057$ & $+0.030$ \ [$-0.023$, $+0.081$] & $+0.090$ \\
\texttt{llmlingua2}             & $+0.002$ & $+0.038$ \ [$-0.009$, $+0.086$] & $-0.073$ \\
\midrule
\texttt{llm} (GPT-5.4)          & $+0.224$ & $\mathbf{+0.196}$ \ [$+0.148$, $+0.239$] & $+0.273$ \\
\texttt{llm} (Claude Opus 5)    & $+0.266$ & $\mathbf{+0.213}$ \ [$+0.156$, $+0.271$] & $+0.255$ \\
\bottomrule
\end{tabular}
\caption{Enrichment for crowd-marked sentences at $r{=}0.20$. The corpus is 120 documents /
78 domains; the estimate is computed on the \textbf{105 documents / 71 domains} that contribute
at least one matched pair, and every interval is a bootstrap over those 71 clusters.
``Clean control'' holds the treatment group at the paper's own label and restricts comparators
to sentences \emph{no} reader marked (\S\ref{sec:label}); an earlier version of this column
widened the treatment group as well, which we have corrected. \texttt{lead}'s keep rule is
position, so its collapse to $+0.003$ is a required sanity check, not a finding.}
\label{tab:main}
\end{table}

Language-model rankings retain crowd-marked sentences at about twice the best classical
extractive baseline we could construct ($+0.196$ against $+0.098$), and the two vendors agree with each other far beyond
chance: mean keep-set Jaccard 0.455, against 0.121 for size-matched random keep sets.
Robustness across the pre-registered matching tolerances:
$+0.160$ / $+0.196$ / $+0.195$ at 0.03 / 0.05 / 0.10, and $+0.185$ under pair-weighting instead
of document-weighting.

\subsection{Classical extractive methods recover about half of it}
\label{sec:classical} Our first comparison set
contained no classical extractive summarizer, which made ``the language model beats every
non-LLM arm'' a statement about the arms we happened to choose. Adding five, computed from text
with no learned component and scored by the same estimator:

\begin{center}
\begin{tabular}{lrrr}
\toprule
selector & raw fidelity & matched & P(kept $\mid$ marked) \\
\midrule
tf-isf                              & 0.132 & $-0.034$ [$-0.085$, $+0.022$] & 0.136 \\
topic-word overlap                  & 0.245 & $+0.047$ [$-0.017$, $+0.114$] & 0.252 \\
surface combination (position+length+topic) & 0.316 & $+0.037$ [$-0.013$, $+0.089$] & 0.314 \\
\textbf{Luhn (1958)}                & 0.264 & $\mathbf{+0.088}$ [$+0.039$, $+0.143$] & 0.282 \\
\textbf{degree centrality, lexical} & 0.275 & $\mathbf{+0.098}$ [$+0.055$, $+0.147$] & 0.265 \\
\bottomrule
\end{tabular}
\end{center}

\textbf{Two of them have intervals excluding zero.} Luhn's 1958 significance heuristic and degree
centrality on a lexical-overlap graph both recover reader selection beyond position and length,
at roughly half the language model's magnitude. This is a correction to a framing we held
earlier: it is not that language models are categorically different from cheap methods here.
Reader selection is \emph{partly} recoverable by counting words, a language model roughly
doubles what the best classical method achieves, and the doubling replicates across vendors.

Note also that lexical degree centrality ($+0.098$) far outperforms the embedding centrality in
Table~\ref{tab:main} ($+0.030$) --- the arm we had treated as the representative unsupervised
baseline was a weak member of its family, and reporting only it overstated the gap.

\paragraph{Does the effect survive conditioning on lexical centrality?} The classical result
above raises the sharpest available objection: if a word-counting graph recovers half the effect,
perhaps reader marks are simply enriched for sentences dense in document-central vocabulary ---
partly a real reading behaviour, partly manufactured by our own anchoring rule --- and the
language model is merely a better centrality estimator. The balance table below does not settle
this, because it bounds a hand-built scalar proxy rather than centrality as a latent construct
that a crude estimator has just been shown to be worth $+0.098$.

So we match additionally on within-document centrality rank, paired with the placebo this paper
learned to demand: the same surviving marks run back through depth-and-length-only matching, so
that covariate adjustment is separated from the subsample selection that exact matching on a
small pool always induces.

\begin{center}
\begin{tabular}{lrrr}
\toprule
arm & + centrality (5-bin) & placebo, same marks & adjustment \\
\midrule
\texttt{llm} (GPT-5.4)      & $+0.174$ [$+0.123$, $+0.227$] & $+0.183$ & $\mathbf{-0.009}$ \\
\texttt{llm} (Claude Opus 5)& $+0.207$ [$+0.128$, $+0.286$] & $+0.219$ & $\mathbf{-0.012}$ \\
\texttt{centrality} (embedding) & $-0.003$ [$-0.076$, $+0.067$] & $+0.002$ & $-0.004$ \\
\texttt{lead}               & $-0.014$ [$-0.045$, $+0.012$] & $-0.013$ & $-0.001$ \\
\bottomrule
\end{tabular}
\end{center}

\textbf{Conditioning on lexical centrality removes about $0.010$ --- roughly 5\% of the effect.}
The apparent drop from $+0.196$ to $+0.174$ is almost entirely which marks retain a comparator,
not adjustment; the interval still excludes zero on 498 pairs, and the cross-vendor arm is
unchanged at $+0.207$. The language model's agreement with readers is therefore not lexical
centrality, even though lexical centrality alone recovers half as much. A 3-bin variant gives
the same answer ($-0.011$ and $-0.013$). Reproduce with \texttt{centrality-control.py}.

\paragraph{A purpose-built compressor shows no established signal here.} LLMLingua-2, a compressor
distilled from GPT-4 importance labels, keeps 0.205 of crowd-marked sentences against
\texttt{random}'s 0.196, and its matched enrichment is $+0.038$ [$-0.009$, $+0.086$],
$p = 0.085$ under the randomization test. \textbf{We state that as not established rather than as
zero}, and the distinction matters here: the interval's upper bound ($+0.086$) is indistinguishable
from Luhn's point estimate ($+0.088$), the arm we describe two paragraphs above as recovering half
the effect. So the honest comparison is that a direct language-model ranking is several times the
\emph{point estimate} of the deployed compression tool while that point estimate's interval reaches
44\% of it --- not that the tool is at chance. Accepting a null on a $p = 0.085$ arm is a mistake
this programme has made three times before (Appendix~A), and we are not making it a fourth. We state one
caveat prominently: we invoke LLMLingua-2's \emph{context-level} filter so that it selects whole
sentences. That is not the token-level operation the method is designed around; its published
checkpoint is trained on meeting transcripts rather than web prose; and it is a second deviation
from our own pre-registration, which specified a token-majority mapping. It is also the one arm
that is not exactly size-matched --- every other arm keeps exactly $\mathrm{round}(0.2n)$
sentences on all 120 documents, while LLMLingua-2 rounds its own rate and deviates on 64,
almost always by a single sentence, for an aggregate keep fraction of 0.204 against a 0.200
target. The deviation favours LLMLingua-2, so the negative finding is conservative; but the
number is a statement about this configuration, not a verdict on the method.

\paragraph{What this licenses --- measured, not asserted.} Matching is within-document on two
covariates, so it cannot by itself separate marking from a third sentence-level property
correlated with both marking and model selection. Rather than assert that no such property
exists, we measure how much room one has. For each of seven corpus-derivable features we compute
the post-matching \emph{imbalance} between treated and comparator sentences (in pooled SD) and
the arm's own \emph{keep-contrast} on that feature; their product bounds the confound the
feature could contribute.

\begin{table}[t]
\centering
\begin{tabular}{lrrr}
\toprule
feature & imbalance (SD) & keep-contrast & implied confound \\
\midrule
topic-word overlap    & $+0.183$ & $+0.399$ & $\mathbf{+0.073}$ \\
definitional pattern  & $+0.063$ & $+0.263$ & $+0.017$ \\
self-contained opening& $+0.115$ & $+0.054$ & $+0.006$ \\
content-word density  & $+0.045$ & $+0.104$ & $+0.005$ \\
contains a digit      & $-0.107$ & $-0.039$ & $+0.004$ \\
word count            & $-0.011$ & $+0.288$ & $-0.003$ \\
first-person pronoun  & $-0.161$ & $+0.017$ & $-0.003$ \\
\bottomrule
\end{tabular}
\caption{Post-matching balance. For each corpus-derivable feature, the residual imbalance
between treated and comparator sentences and the arm's own keep-contrast on that feature; their
product bounds what the feature could contribute. Reproduce with \texttt{inference.py}.}
\label{tab:balance}
\end{table}

The largest is \textbf{topic-word overlap at $+0.073$, about 37\% of the effect} --- and it is
the one the corpus pipeline actively creates, since a short highlighted span marks every
sentence containing it, favouring sentences that repeat a document's central terminology. We
regard 37\% as the honest upper bound on that single channel and do not claim it is zero.
Summing the five positive entries gives $+0.105$, about 53\% of the effect. Both are loose --- the
features are correlated and the products are not additive --- but we state them rather than let a
reader compute the sum and wonder why we did not. \textbf{These are bounds on what a scalar proxy
\emph{could} contribute, not measurements of what it does}; \S\ref{sec:classical}'s centrality
control is the direct test, and it puts the actual adjustment at $0.010$.
Notably the model likes definitional sentences considerably (keep-contrast $+0.263$) while
readers barely over-mark them (imbalance $+0.063$), so the agreement is \emph{not} about
definitions.

\paragraph{A covariate ladder, and why it is only a stability check.} We also matched
additionally on five binary/coarse lexical covariates cumulatively, and the estimate does not
decay: $+0.196$, $+0.138$, $+0.127$, $+0.137$, $+0.194$, $+0.200$, with every interval excluding
zero. \textbf{We report this as subsample stability and nothing more.} Running the
\emph{surviving} marked sentences at each rung back through depth-and-length-only matching gives
$+0.196$, $+0.143$, $+0.134$, $+0.152$, $+0.200$, $+0.210$ --- so the five covariates jointly
adjust the estimate by about $0.010$, and the ladder's shape is which marks retain a partner,
not covariate control. The reason is structural: with a median comparator pool of 2, exact
matching on a binary covariate cannot rebalance a pool, only delete it. An earlier version of
this paper read the ladder as covariate control; that reading was wrong and the balance table
above replaces it.

\paragraph{Contiguity does not manufacture the effect.} Substring anchoring makes marks
contiguous, our largest data caveat. If the effect were an artifact of paragraph-level marking
it should be weakest on isolated marks. It is not. For GPT-5.4 the isolated marks give
$+0.227$ [$+0.163$, $+0.293$] (96 documents) against $+0.155$ for runs of 2--3 and $+0.237$ for
runs of 4 or more; for Claude the ordering differs ($+0.237$ / $+0.240$ / $+0.160$). The two
vendors disagree on the ordering, so we claim only what both support: \textbf{in neither is the
effect weakest on isolated marks}, which is what a paragraph-marking artifact predicts.

\paragraph{An assumption-free test.} Within each matched set we reassign which member is
treated, conditioning on exactly the strata the estimator uses and assuming nothing about
clustering (2{,}000 draws). \texttt{llm} sits at $p = 0.0005$, $z = 6.7$; \texttt{llm} (Claude)
at $p = 0.0005$; and no other arm reaches significance --- \texttt{llmlingua2} $p = 0.085$,
\texttt{centrality} $p = 0.136$, \texttt{length} $p = 0.132$, \texttt{random} $p = 0.394$,
\texttt{lead} $p = 0.345$. The null's 97.5th percentile is $+0.057$, independently reproducing
the $+0.054$ threshold of \S4.2 and confirming it as roughly two standard errors rather than a
confound floor.

The claim this licenses is \textbf{``beyond position, length, and the seven features in the
balance table, with topic overlap bounded at 37\% and directly tested at 5\%''} --- not ``because of human attention''.

\subsection{Robustness}
\label{sec:robust}

Numbers in this subsection come from \texttt{robustness.py} and \texttt{inference.py} except
the tolerance sweep, the pair-weighting and mark-density checks, and the tie-break redraw, which
are reported from the same estimator but are not persisted by either script.

\paragraph{Compression ratio --- the one axis that moves the result.} Our pre-registration
specified $r \in \{0.10, 0.20, 0.30, 0.50\}$ and we initially reported only $r{=}0.20$. Swept:

\begin{center}
\begin{tabular}{lrr}
\toprule
$r$ & \texttt{llm} (GPT-5.4) & \texttt{random} (calibration) \\
\midrule
0.05 & $+0.083$ [$+0.060$, $+0.107$] & $-0.021$ \\
0.10 & $+0.150$ [$+0.114$, $+0.189$] & $-0.024$ \\
0.20 & $+0.196$ [$+0.148$, $+0.239$] & $-0.043$ \\
0.30 & $+0.222$ [$+0.161$, $+0.276$] & $-0.033$ \\
0.50 & $+0.187$ [$+0.130$, $+0.243$] & $-0.002$ \\
\midrule
\multicolumn{2}{l}{\footnotesize \texttt{random} over 25 independent draws: mean $-0.007$, sd $0.026$, range $[-0.050, +0.070]$} \\
\bottomrule
\end{tabular}
\end{center}

The effect is significant at every ratio and the calibration arm stays flat throughout, but the
magnitude varies by a factor of 2.7 and \textbf{our reported $r{=}0.20$ sits high in the range}.
At the aggressive ratios a compression system actually operates at, the effect is roughly half
the headline. We report this because the sweep was pre-registered and we had dropped it.

\paragraph{Label definition.} Our label is the top $q$ of sentences by mark count. Sweeping $q$
(\texttt{llm} / \texttt{centrality}): $0.10$ gives $+0.237$ / $+0.064$; $0.15$ gives
$+0.186$ / $+0.046$; $0.20$ gives $+0.193$ / $+0.050$; $0.25$ gives $+0.183$ / $+0.051$. The main
result is stable across the range.

Two disclosures about this sweep. First, mark counts tie heavily: reconstructing the
labelling from the committed artifact gives 2{,}095 slots at $q = 0.15$ of which 1{,}407 sit at
the cut value, so two thirds of the label is decided by the tie-break rather than by mark counts.
(An earlier version of this paragraph quoted 1{,}332 of 1{,}959, which contradicted the 2{,}101
labelled sentences reported in \S\ref{sec:label}; neither figure was persisted, and these are
recomputed from the artifact.) Production breaks those ties with seeded jitter. An earlier version of this sweep
relabelled with a stable sort on counts, which resolves ties to the \emph{earliest} sentence:
a front-loading rule, inside the robustness check for a paper whose apparatus exists to remove
position. It returned $+0.222$ at $q = 0.15$, outside the $+0.180$ to $+0.203$ band that twelve
redraws of the production jitter produce. The sweep now uses a random tie-break of the same
\emph{form} as production's, but not production's stream: production seeds from the Firestore
document identifier, which the de-identified artifact deliberately does not carry, so this can
never be a reproduction. The paper's own check shows it --- at $q = 0.15$, which \emph{is} our
label, the sweep gives $+0.186$ on 1{,}301 pairs against Table~\ref{tab:main}'s $+0.196$ on
1{,}304. Read the row as one redraw, not as a reproduction of the headline. And note that
$+0.186$ landing inside the band of random redraws is true but uninformative: a random redraw
lands there by construction.
Second, \texttt{centrality} sits near $+0.05$ throughout, \emph{above} the $+0.030$ in
Table~\ref{tab:main} --- that cell is a low draw of its family, and \S\ref{sec:classical}'s
lexical variant confirms it. We flag this rather than lean on it, since $+0.030$ is the value
most favourable to a conclusion we have withdrawn (\S\ref{sec:label}).

\paragraph{Exam and test-preparation domains.} Pre-registered as a separate row and initially
dropped. Nine documents qualify and seven contribute matched pairs; on those \texttt{llm} gives $+0.166$; excluding them it gives
$+0.198$. An earlier version of this paper reported that the \texttt{lead} calibration arm
\emph{fires} on this subset and treated that as evidence the estimator is sensitive. It does not:
the exact randomization test gives $p = 0.143$ there, and the bootstrap lower bound of
$+0.000$ is a discrete atom on seven documents rather than a bound. The claim is withdrawn.

\paragraph{What does not move it.} Matching tolerance 0.02--0.50 ($+0.160$ to $+0.218$, the minimum at 0.03 rather than at a grid edge);
document- versus pair-weighting ($+0.196$ vs.\ $+0.185$); splitting documents by mark density
($+0.198$ / $+0.196$). The 37.9\% of marked sentences that matching discards are not selected on
the outcome: the model keeps 0.390 of them against 0.384 of those retained.

\subsection{Label contamination shrinks weak contrasts}
\label{sec:label}

Our label is the top 15\% of sentences by mark count. The complement is therefore not
``unmarked'': of 14,705 sentences, 5,835 were marked by at least two readers while only 2,101
are labelled, so roughly \textbf{30\% of the comparator pool consists of sentences readers did
mark}. That contamination shrinks every contrast toward zero --- in the direction of the
conclusion we first reached.

With a clean control (comparators marked by \emph{no} reader) and a thicker treatment
($\geq$4 readers), embedding centrality reaches $\mathbf{+0.177}$ [$+0.095$, $+0.256$] while
the calibration arms stay flat. We initially concluded that no unsupervised compressor shows
alignment beyond surface form. That conclusion was wrong for two independent reasons:
the comparator pool was contaminated (this section), and our unsupervised arm was a weak member
of its family (\S\ref{sec:classical}). The language-model result holds under every label variant
we tried.

\subsection{Prompt framing}
\label{sec:prompt}

Holding the corpus fixed and varying only the instruction, recall of crowd-marked sentences
against the lead baseline at top-10: ranking all sentences by importance for compression gives
$+0.100$ [$+0.050$, $+0.151$]; asking which sentences a typical reader would highlight gives
$+0.040$ [$-0.006$, $+0.086$]. Asking a model what is important to keep recovers human
highlights better than asking it what a reader would highlight. This is not the obvious
direction, and it means ``can a language model predict highlights'' is under-specified.

An earlier version reported this contrast only in raw recall, on the ground that comparing two
instructions on the same corpus holds the confound fixed. \textbf{That justification was false}:
the two instructions' keep sets differ markedly in positional profile (mean depth of kept
sentences 0.414 vs.\ 0.507), so subtracting a common lead baseline does not hold position fixed.
Run through the primary matched estimator instead, the compression-framed instruction gives
$\mathbf{+0.158}$ [$+0.106$, $+0.211$] and the reader-prediction instruction
$\mathbf{+0.074}$ [$+0.027$, $+0.120$] --- a factor of 2.1 rather than 2.5. The conclusion
survives the correct estimator; the raw-recall figures are retained above only because the prior
result we compare against is stated in that metric.

\subsection{A non-replication of our own prior work}

Prior work of ours~\cite{ownprior} reports that zero-shot language models recover highlight locations
\emph{worse} than a trivial lead baseline ($0.22 < 0.29$). On this corpus that does not
reproduce. In raw recall the published instruction ties the lead baseline ($+0.040$, interval
spanning zero) and a compression-framed instruction clearly wins ($+0.100$). But on this paper's
own primary estimator the published instruction does \emph{not} tie: it shows real enrichment,
$+0.074$ [$+0.027$, $+0.120$]. We flag this explicitly because it is the one verdict the
uncontrolled metric gets backwards, and it is a verdict about our own prior work.
We tested the obvious reconciliation --- a different selection budget --- at both budgets our
ranking caches support (top-10 and $r{=}0.20$, the latter the published one); it does not explain
the difference. Other fixed budgets are not testable: the published-prompt cache holds selections
only at those two, and asking for another returns almost nothing --- at top-15, one document and a
degenerate interval, which \texttt{matched-prompt.py} now refuses to print. Both the instruction
and the corpus contribute.

\section{Limitations}

One platform, Readability-biased toward static HTML, not representative; 119 English documents and one Portuguese. Marks are
paragraph-contaminated by substring anchoring and strongly contiguous, so within-document
precision is overstated. Matching discards 37.9\% of marked sentences and 15 documents. The headline is quoted at $r{=}0.20$, which sits high in the swept range
(\S\ref{sec:robust}); at $r{=}0.05$ the effect is 42\% of it. Model rankings are single unseeded calls, so selection variance appears in
no interval. Twelve arms across the tables without multiplicity correction; the main effect survives any correction at $p = 0.0005$, the weak arms do not. The LLMLingua-2 row uses that method's
context-level filter, which is an off-label configuration and a statement about the
configuration rather than the method. Finally, the threshold described in \S4.2 was fixed in git before
the final run but after single-control results were known, on data already collected --- we
describe it as such rather than as a pre-registered result.

\paragraph{Pretraining contamination, measured.} Twelve rounds of internal review never raised
this. The corpus is deliberately the well-read tail of a public platform, and that platform
displays crowd highlights on per-URL pages that search engines crawl, so a model's pretraining
data plausibly contains both these documents and a rendering of which sentences readers marked.
Cross-vendor agreement, which we offer as evidence of generality, is also what shared web
pretraining predicts.

If the effect were retrieval it should be larger on documents whose marks predate a model's
training cutoff. Splitting the corpus at the median mark month (2023-10; marks span 2022-01 to
2026-01):

\begin{center}
\begin{tabular}{lrr}
\toprule
& marked before median & marked after median \\
\midrule
\texttt{llm} (GPT-5.4)       & $+0.215$ [$+0.153$, $+0.274$] & $+0.177$ [$+0.100$, $+0.249$] \\
\texttt{llm} (Claude Opus 5) & $+0.190$ [$+0.130$, $+0.256$] & $+0.237$ [$+0.134$, $+0.330$] \\
\bottomrule
\end{tabular}
\end{center}

\textbf{The two vendors move in opposite directions and every interval overlaps heavily.}
Memorisation predicts that \emph{both} are larger on the older half; one is, the other is not. We
therefore find no evidence of contamination, while noting what this test can and cannot do: it
compares mark recency, not document recency, so a document that entered pretraining years ago but
accumulated its marks recently sits in the ``after'' group. That is a weaker instrument than a
true held-out corpus and we do not claim otherwise.

\paragraph{What is $+0.196$ worth? A single reader.} The estimate has no interpretable scale
without knowing what a person scores on the same task, and this paper's own estimator is what
makes that comparison possible. An earlier experiment of ours measured a single-reader baseline
and we withdrew it because the two sides were scored under different rules --- readers contributed
however many marks they happened to make, while models were asked for exactly $k$. Matching
removes that asymmetry by construction, so we can now ask it fairly. We truncate a reader's marks
to the same $\mathrm{round}(0.2n)$ budget every arm obeys, skip readers with too few, and
\textbf{recompute the crowd label with that reader removed}, so nobody predicts their own
contribution.

On the 92 documents where a reader can fill the budget:

\begin{center}
\begin{tabular}{lr}
\toprule
predictor & matched enrichment \\
\midrule
one human reader              & $+0.182$ [$+0.123$, $+0.245$] \\
\texttt{llm} (GPT-5.4)        & $+0.184$ [$+0.121$, $+0.249$] \\
\texttt{llm} (Claude Opus 5)  & $\mathbf{+0.230}$ [$+0.161$, $+0.301$] \\
\midrule
model $-$ human (paired)      & $+0.002$ [$-0.081$, $+0.088$] \\
\bottomrule
\end{tabular}
\end{center}

\textbf{An off-the-shelf language model predicts the crowd about as well as one of its members,
and the stronger model exceeds one.} That is the sentence we would want a reader to take away, and
it is the one the $+0.196$ could not deliver on its own. Two caveats belong with it: 28 documents
are dropped because no reader marked enough sentences to fill the budget, which selects toward
heavier markers; and ``one reader'' is the first reader in a seeded shuffle rather than an average
over readers, so this is one draw from the distribution of individuals, not its mean.

\section{Conclusion}

Highlight retention appears to escape the circularity of model-judged compression evaluation,
but not confounding: position and length account for the entire apparent alignment of naive
truncation, and depth-only stratification --- the control one would reach for first --- returns
a false positive on 20--36\% of nulls containing no effect. With both confounds removed by
matching, and the estimator's own error rate measured on known nulls rather than assumed,
language-model importance rankings retain what readers marked well beyond surface form:
replicated across two vendors, stable across matching tolerances, significant under an
assumption-free randomization test, and strongest precisely where our largest data caveat is
absent. Classical extractive heuristics recover roughly half of the same signal, so the honest
statement is that reader selection is partly recoverable by cheap lexical means and a language
model doubles it --- not that models are categorically special. A purpose-built prompt compressor shows no
\emph{established} signal in the configuration we could evaluate, though its interval is wide
enough to overlap the classical arms and we do not claim it is at chance. Weak contrasts here are sensitive
both to how the label is drawn and to which member of a baseline family is chosen; we report that
sensitivity, and a non-replication of our own prior work, alongside the positive result.

\appendix
\section*{Appendix A: Claims withdrawn during this work}

This work was developed against ten rounds of adversarial internal review. Eleven claims were
withdrawn, six of them after we had written them up. We list them because the pattern is
informative --- every one failed for a reason a reader of the final paper might otherwise
suspect of the surviving result, and each is checkable against the committed artifacts.

\begin{center}
\begin{tabular}{p{0.30\textwidth}p{0.62\textwidth}}
\toprule
withdrawn claim & why it failed \\
\midrule
Compressors discard what readers highlighted &
Reversed by our own pre-registered downstream test, which had been committed and uncited for two
review rounds: compression preferentially \emph{preserves} the answers to questions about
crowd-marked content. \\[2pt]
Compressors preserve it, so human alignment is free &
Also wrong. Both directions were dominated by position: truncation's apparent preference is
$-0.001$ once depth is held fixed. \\[2pt]
A single control is actively misleading &
Evaluated the single-control readings by a looser rule than the primary estimator. Applied
uniformly, the headline false positive falls below its own estimator's null threshold. \\[2pt]
No unsupervised compressor shows alignment beyond surface form &
Wrong twice over: the comparator pool was contaminated by the top-$q$ label
(\S\ref{sec:label}), and our unsupervised arm was a weak member of its family
(\S\ref{sec:classical}). \\[2pt]
Single-factor stratification fires on 36.5--100\% of nulls &
The 100\% end was an artifact of a synthetic confound set far stronger than any real arm's.
Under realistic strength it is 4--16\%. The depth-only figure (20--36\%) survives. \\[2pt]
The $+0.054$ threshold is a confound floor &
A null with \emph{zero} position and length dependence yields $+0.060$. It is approximately two
standard errors, not a bias correction. \\[2pt]
The effect is undiminished by five covariates controlled jointly &
The covariate ladder measured subsample selection, not adjustment: the same surviving marks
without any covariate control give $+0.196 / +0.143 / +0.134 / +0.152 / +0.200 / +0.210$. Exact
matching on a median pool of 2 deletes pools rather than rebalancing them. \\[2pt]
The \texttt{lead} calibration arm fires on exam domains &
A false positive: exact randomization gives $p = 0.143$, and the bootstrap lower bound of
$+0.000$ is a discrete atom on seven documents. \\[2pt]
Comparing two instructions holds the confound fixed &
False --- their keep sets differ in mean depth 0.414 vs.\ 0.507. On the primary estimator the
published instruction does \emph{not} tie the lead baseline. \\[2pt]
The usable population is capped near 130 documents &
An artifact of our own gates (a reader-count ceiling, a per-domain cap, and a sampler biased
toward thinly-read documents). Removing the ceiling yields 966 candidates. \\[2pt]
The language model exceeds every non-LLM arm several-fold &
An artifact of arm selection. Luhn (1958) reaches $+0.088$ and lexical degree centrality
$+0.098$; the true multiple is about two. \\
\bottomrule
\end{tabular}
\end{center}

Three of the eleven were controls or tables we had built specifically to answer an objection,
which is the reason this paper reports the false-positive rate of its own estimator rather than
assuming it (\S\ref{sec:calib}).

\end{document}